\documentclass[a4paper]{article}
\newcommand{\ltwofour}{\frac{\lambda^2}{4}}
\newcommand{\ESch}{\ensuremath{E_{\rm Sch}}}
\newcommand{\ED}{\ensuremath{E_{\rm D}}}
\newcommand{\EKG}{\ensuremath{E_{\rm KG}}}
\newcommand{\Seig}{\ensuremath{\varsigma}}
\newcommand{\Sim}{\mathop{\sim}}
\newcommand{\x}{X}

\newcommand{\lKG}{\ensuremath{\lambda_{{\rm KG}}}}
\newcommand{\lD}{\ensuremath{\lambda_{{\rm D}}}}
\newcommand{\lSch}{\ensuremath{\lambda_{{\rm Sch}}}}
\newcommand{\be}{\begin{equation}}
\newcommand{\eea}{\end{eqnarray}}
\newcommand{\bea}{\begin{eqnarray}}
\newcommand{\ee}{\end{equation}}

\newcommand{\fKG}{\ensuremath{f_{{\rm KG}}}}
\newcommand{\fD}{\ensuremath{f_{{\rm D}}}}
\newcommand{\psiSch}{\ensuremath{\psi_{{\rm Sch}}}}
\newcommand{\psiKG}{\ensuremath{\psi_{{\rm KG}}}}
\newcommand{\psiD}{\ensuremath{\psi_{{\rm D}}}}
\newcommand{\psiDbar}{\ensuremath{\bar{\psi}_{{\rm D}}}}
\newcommand{\e}[1]{\exp({#1})}
\newcommand{\D}{{\rm d}}
\newcommand{\half}{\ensuremath{\frac{1}{2}}}
\newcommand{\AiryAi}{\ensuremath{\mathop{{\rm Ai}}}}
\newcommand{\AiryBi}{\ensuremath{\mathop{{\rm Bi}}}}
\newcommand{\oder}[2]{\frac{\D #1}{\D #2}}
\newcommand{\pder}[2]{\frac{\partial #1}{\partial #2}}
\newcommand{\dt}{\ensuremath{{\rm d}t}}
\newcommand{\dx}{\ensuremath{{\rm d}x}}
\newcommand{\dy}{\ensuremath{{\rm d}y}}
\newcommand{\dz}{\ensuremath{{\rm d}z}}
\newcommand{\ds}{\ensuremath{{\rm d}s}}
\newcommand{\thetahalf}{\frac{\theta}{2}}
\newcommand{\phiprime}{\ensuremath{\check{\phi}}}
\newcommand{\chiprime}{\ensuremath{\check{\chi}}}
\newcommand{\pt}{\ensuremath{\Seig p}} 
\usepackage{amssymb}
\usepackage{amsmath}
\usepackage{cite}
\author{M. Khorrami$^{1,3}$\footnote{mamwad@mailaps.org} , M. Alimohammadi$^{2}$
\footnote{alimohmd@ut.ac.ir}, and A. Shariati$^{1,3}$
\\ $^1$ {\small Institute for Advanced Studies in Basic Sciences,}
\\ {\small P.O.Box 159, Zanjan 45195, Iran.}
\\ $^2$ {\small Department of Physics, University of Tehran,}
\\ {\small North Karegar Ave., Tehran, Iran.}
\\ $^3$ {\small Institute of Applied Physics, P.O.Box 5878, Tehran 15875, Iran.}
} 
\title{Spin $0$ and spin $1/2$ quantum relativistic particles in a constant
gravitaional field
} 
\date{}
\begin{document}
\maketitle
\begin{abstract}

\noindent The Klein-Gordon and Dirac equations in a semi-infinite
lab ($x > 0$), in the background metric $\ds^2 = u^2(x) (-\dt^2 +
\dx^2) + \dy^2 + \dz^2$, are investigated. The resulting equations
are studied for the special case $ u(x) = 1 + g x$. It is shown
that in the case of zero transverse-momentum, the square of the
energy eigenvalues of the spin-$1/2$ particles are less than the
squares of the corresponding eigenvalues of spin-$0$ particles
with same masses, by an amount of $mg\hbar c$. Finally, for
nonzero transverse-momentum, the energy eigenvalues corresponding
to large quantum numbers are obtained, and the results for
spin-$0$ and spin-$1/2$ particles are compared to each other.
\end{abstract}
\section{Introduction}

The behaviour of bosons and fermions in a gravitational field, has
been of interest for many years, from the simplest case of a
nonrelativistic quantum particle in the presence of constant
gravity, \cite{LL} for example, to more complicated cases of the
relativistic spin-$1/2$ particles in a curved space-time with
torsion, \cite{2,2a,3} for example. Several experiments have been
performed to test theoretical predictions, among which are the
experiments of Colella et al. \cite{4}, which detected
gravitational effects by neutron interferometry, and the recent
experiment of Nesvizhevsky et al. \cite{N}, in which the quantum
energy levels of neutrons in the Earth's gravity were detected.

Chandrasekhar considered the Dirac equation in a Kerr-geometry
background, and separated the Dirac equation into radial and
angular parts, \cite{6,7}. In \cite{8}, the angular part was
solved, and in \cite{9} some semi-analytical results for the
radial part were obtained. Similar calculations were performed for
the Kerr-Newman geometry and around dyon black holes in refs.
\cite{10} and \cite{11}, respectively.

In this article, we investigate relativistic spin-$0$ and
spin-$1/2$ particles in a background metric $\ds^2 = u^2(x)
(-\dt^2 + \dx^2) + \dy^2 + \dz^2$, where the particles exist in a
semi-infinite laboratory ($x>0$). The wall $x=0$, which prevents
particles from penetrating to the region $x<0$, corresponds to a
boundary condition. For spin-$0$ particles, this is simply the
vanishing of wave function on the wall. For the spin-$1/2$
particles, it is less trivial and will be discussed in the
article. We consider the Hamiltonian-eigenvalue problems
corresponding to a general function $u(x)$, and obtain the
differential equations and boundary conditions corresponding to
the spin-$0$ and spin-$1/2$ particles. The special case
$u(x)=1+gx$, is investigated in more detail. An exact relation
between the square of the energy eigenvalues of spin-$0$ and
spin-$1/2$ particles, whit same masses and no transverse momenta,
is obtained, namely $\ED^2 = \EKG^2 - mg\hbar c$. Finally, the
energy eigenvalues for large energies are obtained.

\section{Review of the non-relativistic problem}
The potential energy of a non-relativistic particle in a constant
gravitational field is $V_{{\rm grav}}(x) = mgx$, where $g$ is the
acceleration of gravity, the direction of which is along the $x$
axis, towards the negative values of $x$.

The Schr\"{o}dinger equation is written as $H \psiSch = i \hbar
\partial_t \psiSch $, where $H=\mathbf{P}^2/(2m)+V_{{\rm grav}}$,
subject to the boundary conditions that $\psiSch$ does not
diverge at ${x\to \pm\infty}$.

Writing $\psiSch = \exp[(-i E t + i p_2 y + i p_3 z)/\hbar]F(x)$,
it is easily seen that $F(x)$ must satisfy $F''(x)=(L^{-3}x
+L^{-2}\lSch)F(x)$, where a prime means differentiation with
respect to the argument, $L^{-2}\lSch :=(p_2^2+p_3^2-2 m
E)/(\hbar^2)$, and $L:=(2 m^2 g/\hbar^2)^{-1/3}$. This equation
has two linearly independent solutions: the Airy functions
$\AiryAi(L^{-1}x+\lSch)$ and $\AiryBi(L^{-1}x+\lSch)$ (see for
example p. 569 of \cite{BO}). \AiryBi\ violates the boundary
condition at $+\infty$. So the solutions are
$\AiryAi(L^{-1}x+\lSch)$. These functions don't tend to zero at
$x\to -\infty$, and it is expected, since the potential $mgx$ is
not bounded from below. But we note that a lab usually has walls.
Consider a semi-infinite lab, for which
\begin{equation}
\label{wall} V_{{\rm walls}} = \left\{
      \begin{array}{ll}
        + \infty & x < 0 \cr
        0        & x \geq 0.
      \end{array}
\right.
\end{equation}
In such a lab, the boundary condition at $-\infty$ is replaced
with
$ 
\lim_{x\to 0^+} \psiSch = 0.
$ 
So the solutions are $\AiryAi(L^{-1}x + \lSch)$, where $\lSch$
must be one of the zeros of \AiryAi.

The first four zeros of \AiryAi\ are approximately $\lambda_1 =
-2.3381$, $\lambda_2 = -4.0879$, $\lambda_3 = -5.5206$, and
$\lambda_4 = -6.7867$.  If the transverse momenta of the particle
vanish, the energy levels are these numbers multiplied by
$-\hbar^{2/3} (2m)^{1/3} (g/2)^{2/3}$, the value of which for a
neutron, in a lab on the Earth, where $g \cong 9.8 \, {\rm m} {\rm
s}^{-2}$, is $0.59 \, {\rm peV}$. Therefore, the first 4 energy
levels of a neutron are $E_1 = 1.4 \, {\rm peV}$, $E_2 = 2.5 \,
{\rm peV}$, $E_3 = 3.3 \, {\rm peV}$, and $E_4 = 4.0 \, {\rm
peV}$. This result has been recently verified experimentally
\cite{N}.

\section{A relativistic quantum particle}
According to general relativity, gravity is represented by a
pseudo-Riemannian metric $\ds^2 = g_{\mu\nu} \dx^\mu \dx^\nu$. We
use the signature $(- + + +)$ for the metric. The form of the
metric depends on both the gravitational field and the coordinate
system used to describe the field.

In a spacetime with metric $g_{\mu\nu}$, the Klein-Gordon
equation, the equation for a spinless massive particle, is
\begin{equation}
\left( \frac{1}{\sqrt{-g}} \pder{}{x^\mu} \sqrt{-g} \, g^{\mu\nu}
\pder{}{x^\nu} - m^2 \right) \psiKG = 0,\nonumber
\end{equation}
where $g := \det \left[g_{\mu\nu}\right]$. We have used a system
of units in which the numerical values of the velocity of light
($c$) and the Planck constant (divided by $2\pi$) are both unity.

To write the Dirac equation in a curved spacetime, or a flat
spacetime but in a curvilinear coordinate system, one can employ
the Equivalence Principle~--~see for example pp. 365--373 of
\cite{W}, but note that our convention for the Dirac Lagrangian,
and therefore the Dirac equation, is different from that of
Weinberg: we use `$-m$' in the following equation, instead of a
`$+m$' in Weinberg's. It reads as follows:
\begin{equation}
\gamma^a \left(\partial_a + \Gamma_a \right) \psiD - m \psiD =
0,\nonumber
\end{equation}
where $\Gamma_a$s are spin connections, obtained from the dual
tetrad $e^a$, through
\begin{align}
\D e^a + \Gamma^a{}_b \wedge e^b =& 0\nonumber\\
\Gamma^a{}_b :=& \Gamma^a{}_{cb} \, e^c\nonumber\\
\Gamma_a :=& \half S_{bc} \, \Gamma^c{}_a{}^b\nonumber\\
S_{bc} :=& - \frac{1}{4} \left[ \gamma_b, \gamma_c \right].
\nonumber
\end{align}

We consider a gravitational field which is represented by the
metric
\begin{equation} \label{A}
\ds^2 = u^2(x) \left( - \dt^2 + \dx^2 \right) + \dy^2 + \dz^2.
\end{equation}
One can write the above metric, also like
\begin{equation}
\ds^2 = -U^2(X) \dt^2 + \D X^2 + \dy^2 + \dz^2,\nonumber
\end{equation}
where $(\D X)/(\D x)=U(X)=u(x)$. In a small region of space
(compared to the length in which the gravitaional field changes
significantly) one can use $u(x) = 1 + g x$, $U(X) = 1 + g X$, and
$X=x$. We will investigate the case of a general $u(x)$, but then
limit ourselves to the special case $u(x)=1+gx$.

For the metric (\ref{A}), the Klein-Gordon equation reads
\begin{equation}
\left[ \frac{1}{u^2}  \left( -\pder{^2}{t^2} +\pder{^2}{x^2}
\right) +\pder{^2}{y^2} + \pder{^2}{z^2} - m^2 \right] \psiKG =
0.\nonumber
\end{equation}

To write the Dirac equation, we need the spin connections. The
only non-vanishing spin connections for metric (\ref{A}) are $
\Gamma^0{}_1 = \Gamma^1{}_0 = (u'/u^2) e^0$. Thus, $
\Gamma^0{}_0{}^1 = - \Gamma^1{}_0{}^0 = u'/u^2,$ and $\Gamma_0 =
\gamma_0 \gamma_1 u'/(2u^2).$ Therefore, the Dirac equation for
this metric reads as
$ 
\left( \gamma^a \partial_a + \half \gamma_1
\displaystyle{\frac{u'}{u^2}} - m \right) \psiD
= 0,
$ 
which means
\begin{equation} \label{DE}
\left[
     \frac{1}{u}
     \left( \gamma^0 \pder{}{t} + \gamma^1 \pder{}{x} \right)
   + \half \gamma^1 \frac{u'}{u^2}
   + \gamma^2 \pder{}{y}
   + \gamma^3 \pder{}{z}
   - m
\right] \psiD = 0.
\end{equation}

\section{Boundary condition at the infinite barrier}
In the non-relativistic equation, the infinite potential barrier
which prevents particles from penetrating the region $x<0$, leads
to the boundary condition $ \lim_{x \to 0^+} \psiSch = 0$. This
boundary condition emerges from the fact that the Schr\"{o}dinger
equation is second order in $x$, from which it follows that
$\psiSch$ must be continuous at $x=0$.

The Klein-Gordon equation is also second order in $x$, so the same
boundary condition
\begin{equation} \label{KGBC}
\psiKG(0) = 0
\end{equation}
emerges. But the Dirac equation is of \textit{first} order, so the
four-spinor $\psiD$ can be discontinuous at $x=0$, if the
potential goes to infinity there. To find the proper boundary
condition, one must first find a proper way of confining a Dirac
particle to the region $x>0$. The first guess is to add a step
potential to the Hamiltonian. But this is the same as adding an
electrostatic potential, which is the time component of a
four-vector, having opposite effects on particles and
anti-particles. Such a potential will not result in a wave
function decaying as $x\to -\infty$. A better way is to add a term
$- V \psiDbar \psiD$ to the Lagrangian, where $V$ is a function of
$t$, $x$, $y$, and $z$~--~something like a Higgs term.  As a
result of adding this `scalar' potential to the Lagrangian, the
mass $m$ of the Dirac praticle is replaced by $m +  V$. So the
Dirac equation reads
\begin{equation} \left( \gamma^\mu
\pder{}{x^\mu} - m - V \right) \psiD = 0.\nonumber
\end{equation}
Now let's consider the step potential
\begin{equation}
\label{V} V(x) =
\begin{cases}
    V, &  x < 0 \\
    0, &  x \geq 0
\end{cases}
\end{equation}
where $V$ is a positive constant. If the energy and the transverse
momenta are finite, and the constant $V$ is large, one can neglect
the parts containing derivatives with respect to $t$, $y$, and
$z$, arriving at
\begin{equation} 
\left( \gamma^1 \partial_x - m  - V(x) \right) \psiD(x) =
0.\nonumber
\end{equation} 
The solution to this equation, which does not diverge as $x\to
-\infty$, is $\psiD \propto \e{qx}$ for $x <0$, where $q$ is a
positive constant. As this equation is of first order, the
solution must be continuous and we must have
\begin{equation}
\left( \gamma^1 q - m - V \right) \psiD(0^+) = 0 \hskip 0.5em
\Rightarrow \hskip 0.5em q^2 = (m + V)^2 \hskip 0.5em \Rightarrow
\hskip 0.5em q = m + V \hskip 1em (q
> 0).\nonumber
\end{equation}
This means that
\begin{equation} \label{bcond}
\left(\gamma^1 - 1 \right) \psiD(0) = 0,
\end{equation}
which must be true, also for $V\to \infty$. Therefore, instead of
(\ref{KGBC}), (\ref{bcond}) is the boundary condition for a Dirac
particle at the floor of the lab.

\section{The Klein-Gordon equation}
Since $u$ in metric (\ref{A}) does not depend on $t$, $y$, and
$z$, one can seek a solution whose functional form is as
$\psiKG(t,x,y,z) = \exp(-i E t + i p_2 y + i p_3 z) \psiKG(x)$. We
arrive at
\begin{equation}
\left[ E^2 + \oder{^2}{x^2} - \left(p^2 + m^2 \right) u^2(x)
\right] \psiKG(x) = 0, \hskip 2em x > 0,
\end{equation}
subject to the boundary condition $\psiKG(0) = 0$, where
$p:=\sqrt{(p_2)^2+(p_3)^2}$. Defining
\begin{equation}\label{xi}
\varepsilon := \sqrt{p^2 + m^2}, \hskip 1em \xi:=\varepsilon x,
\end{equation}
we get
\begin{equation}\label{KGEQ}
\left[ \left( -\oder{}{\xi} + u \right) \left( \oder{}{\xi} + u
\right) + \oder{u}{\xi} \right]
\psiKG\left(\frac{\xi}{\varepsilon}\right) =
\left(\frac{E}{\varepsilon} \right)^2
\psiKG\left(\frac{\xi}{\varepsilon}\right).
\end{equation}

\section{The Dirac equation}
The functions in the equation (\ref{DE}), do not depend on $t$,
$y$, and $z$; therefore energy $E$ and transverse momenta $p_2$
and $p_3$ are constants. By a suitable choice of coordinates (and
without loss of generality) one can set $p_3 =0$, and $p_2 = p$.
So we seek the solution as $\psiD(t, x, y, z)  = \exp(- i E t + i
p y) \psiD(x)$. Inserting this ansatz in (\ref{DE}), the resulting
equation in terms of the $\alpha$ and $\beta$ matrices becomes
\begin{equation}\label{dea}
\left(     \frac{E}{u} \beta
   + \frac{i}{u} \beta \alpha_1 \oder{}{x}
   + \frac{i}{2}  \beta \alpha_1 \frac{u'}{u^2}
   - p \beta \alpha_2
   - m
\right) \psiD = 0.
\end{equation}
Defining
\begin{equation}
\hat{\psi} := \sqrt{u} \psiD,\nonumber
\end{equation}
one arrives at
\begin{equation} \label{B}
\left( E + i \alpha_1 \oder{}{x} - u p \alpha_2  - m u \beta \right)
\hat{\psi} = 0.
\end{equation}
In terms of the so called long and short spinors
\begin{equation}
\phi := \half \left(1+\beta\right) \hat{\psi}, \hskip 2em
\tilde{\chi}:= \half \left(1 - \beta \right) \hat{\psi},
\end{equation}
which are eigenspinors of $\beta$ with eigenvalues $+1$ and $-1$,
respectively, (\ref{B}) becomes
\begin{align}
\left( E - m u \right) \phi + \left( i \alpha_1 \oder{}{x} - u p
\alpha_2 \right) \tilde{\chi} &=0,\nonumber\\
\left( i \alpha_1\oder{}{x} - u p \alpha_2 \right) \phi + \left( E
+ m u \right) \tilde{\chi} &=0.\nonumber
\end{align}
Defining further $\chi := \alpha_1 \tilde{\chi}$, which satisfies
$\beta \chi = \chi$, the above equations read as
\begin{align} \label{C1}
\left( E - m u \right) \phi + \left( i \oder{}{x} - u p \alpha_2
\alpha_1 \right) \chi & = 0,
\\ \label{C2}
\left( i \oder{}{x} - u p \alpha_1 \alpha_2 \right) \phi + \left(
E + m u \right) \chi & = 0.
\end{align}
The matrix $S$ defined through
\begin{equation} \label{S} S:= - i
\beta \alpha_1 \alpha_2,
\end{equation}
commutes with $\alpha_1$, $\alpha_2$, and $\beta$; and its
eigenvalues are $\pm 1$. So one can take $\psiD$, and hence
$\hat\psi$, $\phi$, $\tilde\chi$, and $\chi$, eigenspinors of $S$
with the same eigenvalue $\Seig$, in terms of which equations
(\ref{C1}) and (\ref{C2}) can be written as
\begin{equation}
\left[ \begin{matrix}
   \left( E - m u \right) & i \left(\oder{}{x} +\pt u\right) \\
   i \left( \oder{}{x} - \pt u\right) & \left( E + m u \right) \end{matrix}
\right] \left[
\begin{matrix}
\phi \\ \chi
\end{matrix}
\right] = 0,
\end{equation}
or
\begin{equation} \label{Z}
E \left[
\begin{matrix}
\phi \\ \chi
\end{matrix}
\right] = \left( m u \sigma_3 + \pt u \sigma_2 - i \sigma_1
\oder{}{x} \right) \left[
\begin{matrix} \phi \\ \chi
\end{matrix}
\right],
\end{equation}
where $\sigma$s are the usual Pauli matrices. Defining
\begin{align}\label{theta}
\varepsilon :=& \sqrt{p^2 + m^2} \, ,\nonumber\\
\theta :=& \arctan \frac{\pt}{m} \, , \hskip 2em
          -\frac{\pi}{2} < \theta < \frac{\pi}{2},
\end{align}
we have
\begin{equation}
m \sigma_3 + \pt \sigma_2 = \varepsilon \left( \sigma_3 \cos\theta
+ \sigma_2 \sin\theta \right) = \varepsilon
\exp\left(\frac{i}{2}\sigma_1 \theta\right) \sigma_3
\exp\left(-\frac{i}{2} \sigma_1 \theta\right).\nonumber
\end{equation}
Defining
\begin{equation} \label{bar}
\left[
\begin{matrix}
\phiprime \\ \chiprime
\end{matrix}
\right]  := \exp\left(-\frac{i}{2}\sigma_1\theta\right) \left[
\begin{matrix}
\phi \\ \chi
\end{matrix}
\right] = \left[
\begin{matrix}
      \phi\cos\thetahalf - i \chi\sin\thetahalf \\
          - i \phi  \sin\thetahalf + \chi\cos\thetahalf
\end{matrix}
            \right],
\end{equation}
one gets
\begin{equation}
E \left[
\begin{matrix}
\phiprime \\ \chiprime
\end{matrix}
\right] = \left( \varepsilon u \sigma_3 - i \sigma_1 \oder{}{x}
\right) \left[
\begin{matrix}
\phiprime \\ \chiprime
\end{matrix}
\right],
\end{equation}
which can be written as
\begin{equation}
\left[
\begin{matrix}
E - \varepsilon u & i \displaystyle{\oder{}{x}} \\
   i \displaystyle{\oder{}{x}} & E + \varepsilon u
\end{matrix}
\right] \left[
\begin{matrix}
\phiprime \\ \chiprime
\end{matrix}
\right] = 0.\nonumber
\end{equation}
Introducing $\phi^\pm := \phiprime \pm i \chiprime$, this equation
is transformed to
\begin{equation} \label{EQ}
\left[
\begin{matrix}
E & - \displaystyle{\oder{}{x}} - \varepsilon u \\
\displaystyle{\oder{}{x}} - \varepsilon u & E
\end{matrix}
\right] \left[
\begin{matrix}
\phi^+ \\
\phi^-
\end{matrix}
\right] = 0,
\end{equation}
which, in terms of the variable $\xi = \varepsilon x$,
leads to the following second order differential equation for $\phi^-$
\begin{equation} \label{2ODE}
\left(-\oder{}{\xi} + u \right) \left(\oder{}{\xi} + u \right)
\phi^-\left(\frac{\xi}{\varepsilon}\right) =
\left(\frac{E}{\varepsilon}\right)^2 \phi^-\left(\frac{\xi}
{\varepsilon}\right).
\end{equation}
$\phi^+$ can be obtained from $\phi^-$ by
\begin{equation} \label{19}
\phi^+\left(\frac{\xi}{\varepsilon}\right)=\frac{\varepsilon}{E}
\left(\oder{}{\xi} + u
\right)\phi^-\left(\frac{\xi}{\varepsilon}\right).
\end{equation}

Now turn to the boundary condition. In terms of $\phi$ and $\chi$,
eq. (\ref{bcond}) is written as
\begin{equation} \label{Bcond}
\left( \phi - i \chi \right) \vert_{x=0} = 0.
\end{equation}
Using (\ref{bar}), we have
\begin{equation}
\phi - i \chi =\phi^- \cos\thetahalf + \phi^+ \sin\thetahalf,
\nonumber
\end{equation} from which and (\ref{19}), the boundary condition
for $\phi^-$ follows.
\begin{equation}\label{2OBC}
\phi^-(0)\cos\thetahalf + \frac{\varepsilon}{E} \left(
\oder{}{\xi} + u \right) \phi^-(0) \sin\thetahalf = 0.
\end{equation}
What remains, is to solve equation (\ref{2ODE}) subject to the
boundary condition (\ref{2OBC}).

\section{The special case $u(x) = 1 + g x$}
\subsection{The eigenfunctions}
For the special case $u(x) = 1 + g x$, we introduce
\begin{equation}\label{X}
X := \sqrt{\frac{g}{\varepsilon}} \xi +
\sqrt{\frac{\varepsilon}{g}},
\end{equation}
and the Klein-Gordon equation (\ref{KGEQ}) becomes
\begin{equation} \label{KGfeq}
\left[
 \left( -\oder{}{X} + X \right) \left( \oder{}{X} + X \right) + 1
\right] \fKG(X) = \lKG^2 \fKG(X),
\hskip 2em X \geq \sqrt{\frac{\varepsilon}{g}},
\end{equation}
where
$$ \lKG^2 := \frac{\EKG^2}{g\varepsilon}, $$
and
$$ \fKG(X) := \psiKG(x). $$
The boundary condition reads
(see equation (\ref{KGBC})) \be \label{KGfbc}
\fKG\left(\sqrt{\frac{\varepsilon}{g}}\right) = 0. \ee

For the Dirac equation, in this special case, again in terms of
the variable $X$, equation (\ref{2ODE}) becomes
\begin{equation} \label{Dfeq}
\left( -\oder{}{X} + X \right) \left( \oder{}{X} + X \right) \fD(X) =
\lD^2 \fD(X),
\hskip 2em X \geq \sqrt{\frac{\varepsilon}{g}},
\end{equation}
where
$$ \lD^2 := \frac{\ED^2}{g\varepsilon}, $$
and
$$\fD(X) := \phi^-(x). $$
The boundary condition (\ref{2OBC}), is now
\begin{equation}\label{Dfbc}
\fD\left(\sqrt{\frac{\varepsilon}{g}}\right) \cos\thetahalf +
\frac{1}{\lD} \left( \oder{}{X} + X \right)
\fD\left(\sqrt{\frac{\varepsilon}{g}}\right) \sin\thetahalf = 0.
\end{equation}

Comparing  (\ref{KGfeq}) and (\ref{Dfeq}), and defining $\lambda^2
= \lKG^2 - 1$ for the Klein-Gordon equation and $\lambda^2 =
\lD^2$ for the Dirac equation, we see that both of them are of the
same form:
\begin{equation} \label{eq}
\left( - \oder{}{\x} + \x \right) \left( \oder{}{\x} + \x \right)
f(\x) = \lambda^2 f(\x),
\end{equation}
but subject to different boundary conditions (\ref{KGfbc}) and
(\ref{Dfbc}).

Note that this final equation is the same as the Schr\"{o}dinger
equation for a one-dimensional simple harmonic oscillator.  The
difference is that for the simple harmonic oscillator, $-\infty <
X < +\infty$; while in our case $ \sqrt{\varepsilon/g} < X <
+\infty$, and that we have the boundary condition (\ref{KGfbc}) or
(\ref{Dfbc}) for $X \to \sqrt{\varepsilon/g}^+$.

To solve (\ref{eq}), we define $h(X)$ as
\begin{equation}\label{hasan}
f(\x) = h(\x) \e{-\half \x^2},
\end{equation}
and obtain
\begin{equation}\label{heq}
\left( 2 \x - \oder{}{\x} \right) \oder{h}{\x} = \lambda^2 h.
\end{equation}
Now we seek a power-series solution for $h(X)$:
$$ h(\x) =\sum_{0}^{\infty} a_n \x^n.$$
Putting this in (\ref{heq}), one obtains
\begin{align}
a_{2k}&= \frac{4^k \, \Gamma\left(k - \ltwofour \right)}
{\left( 2k \right)! \, \Gamma \left( - \ltwofour \right)} a_{0},\\
a_{2k+1}&=  \frac{4^k \, \Gamma\left(k +\frac{1}{2}-\ltwofour
\right)} {\left( 2k+1 \right)! \, \Gamma \left(\frac{1}{2}-
\ltwofour \right)} a_{1},
\end{align}
and from that the following two soultions.
\begin{align}
h_0(\x)&= \frac{a_0}{\Gamma\left( -\ltwofour \right)}
\sum_{k=0}^{\infty} \frac{(2\x)^{2k} \Gamma\left(k -
\ltwofour\right)}{(2k)!}, \\
h_1 (\x)&= \frac{a_1}{2\Gamma\left(\frac{1}{2}-\ltwofour \right)}
\sum_{k=0}^{\infty} \frac{(2\x)^{2k+1} \Gamma\left(k+\frac{1}{2}-
\ltwofour\right)} {(2k+1)!}.
\end{align}
We have to find a linear combination of these two functions, which
remains finite as $\x\to\infty$. To do so, we define the function
$$ S_\alpha (\x) := \sum_{k=0}^{\infty}
\frac{(2\x)^{2k+\alpha} \Gamma\left(k+\frac{\alpha}{2} - \ltwofour
\right)} {(2k+\alpha)!}, $$ It is seen that $h_0$ and $h_1$ are
proportional to $S_0$ and $S_1$, respectively. Using a
steepest-descent analysis to obtain the large-$X$ behavior of
$S_\alpha$, it is seen that this behavior is in fact independent
of $\alpha$. This comes from the fact that if one replaces the
summation over $k$ with an integration, and change the variable
$k+\alpha$ into $k$, then the dependence of $S$ on $\alpha$ comes
solely from the lower bound of the integration region. But this
bound is unimportant, since the major part of the sum comes from
large $k$'s. In fact, one can perform the steepest descent
analysis and find
$$ S_\alpha(\x) \Sim\limits_{\x\to\infty} \x^{-1 -\frac{\lambda^2}{2}} \,
\e{\x^2 + 1 + \ltwofour}. $$ So, $h_0(\x) + h_1(\x)$ remains
finite as $\x\to\infty$, provided
$$ \frac{a_1}{2\Gamma\left(\frac{1}{2} -\ltwofour \right)} = -
\frac{a_0}{\Gamma\left( -\ltwofour \right)} =: b. $$

So the unique (up to normalization) normalizable function which
solves (\ref{eq}), is
\begin{equation}\label{h}
 h(\x) := b \sum_{n=0}^{\infty}\frac{(-2\x)^n \,
 \Gamma\left(\frac{n}{2} - \ltwofour \right)}{n!}.
\end{equation}

\subsection{Comparing the energy eigenvalues}
To compare the energies of three different Hamiltonians, namely
the Schr\"{o}dinger equation, the Klein-Gordon equation
(\ref{KGfeq}), and the Dirac equation (\ref{Dfeq}), we reintroduce
the physical constants previously chosen to be equal to $1$. One
obtains
\begin{align}
\ESch &=\frac{p^2}{2m} - 2^{-1/3} \, \lSch \left(\frac{\hbar
g}{mc^3}\right)^{2/3} mc^2,\nonumber\\
\EKG &=\lKG \left(\frac{\hbar
g}{mc^3}\right)^{1/2}mc^2\left(1+\frac{p^2}{m^2c^2}\right)
^{1/4},\nonumber\\
\ED &=\lD \, \left(\frac{\hbar
g}{mc^3}\right)^{1/2}mc^2\left(1+\frac{p^2}{m^2c^2}\right)
^{1/4}.\nonumber
\end{align}

To compare \lD\ and \lKG, we first note that if $p = 0$, then
$\theta = 0$ (see equation (\ref{theta})). Therefore, the two
boundary conditions (\ref{KGfbc}) and (\ref{Dfbc}) become the
same, and eigenvalue $\lambda^2$ in (\ref{eq}), becomes the same
for the Klein-Gordon and the Dirac case. But $\lambda^2$ is equal
to $\lKG^2 - 1$, and $\lD^2$. So,
$$ \lD^2 = \lKG^2 - 1, \hskip 1em \mbox{for} \hskip 0.5em p = 0,
$$
or
\begin{equation} \label{taqi}
   \ED^2 = \EKG^2 -  m\, g\,
\hbar\, c, \hskip 1em \mbox{for} \hskip 0.5em p = 0.
\end{equation}
This is an exact result which determines the effect of spin on the
gravitational interaction of relativistic particles. A fermion and
a boson with same masses, have different quantum energies when
fall vertically in a constant gravitational field.

For $p \neq 0$, the relation between the eigenvalues is more
complicated, because although the differential equations are the
same, the boundary conditions are different. In this case, it is
possible to find approximate relations for $\lKG$ and $\lD$, for
large values of them, and compare them in this region.

A WKB analysis of the wave function $h(X)$, performed in appendix
A, shows that
\begin{equation} \label{akbar}
 h(\x) \Sim\limits_{\lambda\to\infty} \cos
\left( \pi\ltwofour - \lambda \x \right).
\end{equation}
Inserting (\ref{hasan}) in (\ref{Dfbc}), the boundary condition on
$h(\x)$ for Dirac particles reads
$$\left[ h(\x) \, \cos
\frac{\theta}{2} + \frac{1}{\lD} \, \oder{h}{\x} \, \sin
\frac{\theta}{2} \right]\Big\vert_{\x=\sqrt{\varepsilon/g}} = 0,
$$
which, using (\ref{akbar}), leads to
$$\cos \left(
\frac{\pi\lD^2}{4} - \lD \x - \frac{\theta}{2}
\right)\Big\vert_{\x=\sqrt{\varepsilon/g}} = 0, $$ or
\begin{equation}\label{41}
 \frac{\pi\lD^2}{4} - \lD
\sqrt{\frac{\varepsilon}{g}} - \frac{\theta}{2} = \left( n + \half
\right) \pi.
\end{equation}

For the Klien-Gordon equation, from (\ref{KGfbc}) and
(\ref{hasan}), it is seen that $h(\sqrt{\varepsilon/g})=0$. Using
(\ref{akbar}), and remembering $\lambda^2=\lKG^2-1$, we find
\begin{equation}\label{42}
\frac{\pi\lKG^2}{4} - \frac{\pi}{4} - \lKG
\sqrt{\frac{\varepsilon}{g}} = \left( n + \half \right) \pi,
\end{equation}
in which O($1/\lKG$) terms have been ignored. Comparing (\ref{41})
and (\ref{42}), results in
$$
 \lD^2 = \lKG^2 - 1 + \frac{2\theta}{\pi},
$$
or
\begin{equation} \ED^2 = \EKG^2 -
\sqrt{p^2 + m^2 c^2} \, \hbar \, g \left( 1 - \frac{2}{\pi}
\arctan \frac{\Seig p c}{m} \right).
\end{equation}

\appendix
\section{The asymptotic behaviour of the function $h(\x)$}
To obtain the asymptotic behaviour of $h(X)$, eq.(\ref{h}), for
$\lambda\gg 1$ and finite $X$, we first write (\ref{h}) as
\begin{equation}\label{ah}
 h(\x) = b \sum_{k=0}^{\infty}
\frac{(2\x)^{2k} \, \Gamma\left( k - \ltwofour \right)}{(2k)!} -b
\sum_{k=0}^{\infty} \frac{(2\x)^{2k+1} \, \Gamma\left(
k+\frac{1}{2} - \ltwofour \right)}{(2k+1)!}.
\end{equation}
Using $\Gamma (p)\Gamma (1-p)=\pi/(\sin p\pi)$, we have
\begin{equation}\label{A2}
\Gamma\left( k - \ltwofour \right)=\frac{\pi}{\sin\left[\pi (k-
\ltwofour )\right]\Gamma\left(\ltwofour
-k+1\right)}=-\frac{(-1)^k\pi}{\sin\left(\pi\ltwofour
\right)\Gamma\left(\ltwofour -k+1\right)}.
\end{equation}
As $X$ is finite, large $k$'s have negligible contributions in the
power series for $h(X)$. So we can use the Stirling's formula
$\Gamma (x+1)\sim\sqrt{2\pi}x^{x+(1/2)}e^{-x}$, to obtain
\begin{align}
\Gamma\left(\ltwofour -k+1 \right)=&\sqrt{2\pi}\left(\ltwofour
\right)^{\ltwofour-k+\frac{1}{2}} \left( 1-\frac{k}{\lambda^2/4}
\right)^{\ltwofour-k+\frac{1}{2}}e^{k-(\lambda^2/4)}\nonumber\\
&\sim
\sqrt{2\pi}\left(\ltwofour\right)^{\ltwofour-k+\frac{1}{2}}e^{-\lambda^2/4}.
\nonumber
\end{align}
Therefore (\ref{A2}) leads to
\begin{equation}\label{A3}
\Gamma\left( k - \ltwofour \right)\sim
-\frac{(-1)^k\pi}{\sqrt{2\pi} \sin \left( \pi \ltwofour \right) }
\left( \ltwofour \right)^{k-(1/2)}
\left(\frac{4e}{\lambda^2}\right)^{\lambda^2/4}.
\end{equation}
A similar argument shows that
\begin{equation}\label{A4}
\Gamma\left( k +\frac{1}{2}- \ltwofour \right)\sim\frac{(-1)^k\pi}
{\sqrt{2\pi} \cos \left( \pi \ltwofour \right) }\left( \ltwofour
\right)^k \left(\frac{4e}{\lambda^2} \right)^{\lambda^2/4}.
\end{equation}
Inserting (\ref{A3}) and (\ref{A4}) in (\ref{ah}), one obtains
\begin{align}
h(\x) \sim& -b\sqrt{\frac{\pi}{2}} \left( \ltwofour \right)^{-1/2}
\left(\frac{4e}{\lambda^2}\right)^{\lambda^2/4}\nonumber\\
& \times\left[\frac{1}{\sin \left( \pi \ltwofour \right)}
\sum_{k=0}^{\infty}\frac{(-1)^k (\lambda\x)^{2k}}{(2k)!}+
\frac{1}{\cos \left( \pi \ltwofour \right)}
\sum_{k=0}^{\infty}\frac{(-1)^k (\lambda\x)^{2k +1}}{(2k+1)!}
\right]\nonumber\\
=& -\frac{2b\sqrt{2\pi}}{ \lambda \sin
\left(\pi\frac{\lambda^2}{2} \right)} \left(\frac{4e}{\lambda^2}
\right)^{\lambda^2/4} \cos \left( \pi \ltwofour - \lambda
\x\right).
\end{align}
This is nothing, but the leading-order result of a WKB analysis.

\end{document}